\documentclass[%
 aip,
 amsmath,amssymb,
 reprint,
]{revtex4-1}

\usepackage{graphicx}
\usepackage{dcolumn}
\usepackage{bm}
\usepackage[utf8]{inputenc}
\usepackage[T1]{fontenc}
\usepackage{mathptmx}
\usepackage{etoolbox}
\usepackage{hyperref}
\usepackage[english]{babel}
\usepackage{amsfonts}
\usepackage{csquotes}
\usepackage{caption}
\usepackage{gensymb}

\DeclareCaptionLabelFormat{mylabel}{#1 #2.\hspace{1.5ex}}
\makeatletter
\def\@email#1#2{%
 \endgroup
 \patchcmd{\titleblock@produce}
  {\frontmatter@RRAPformat}
  {\frontmatter@RRAPformat{\produce@RRAP{*#1\href{mailto:#2}{#2}}}\frontmatter@RRAPformat}
  {}{}
}%
\makeatother

\begin{document}

\preprint{AIP/123-QED}

\title{Residual stress gradient in a thin film within the dislocation pile-up theory}

\author{A. V. Druzhinin}
 \affiliation{Ioffe Institute, Polytekhnicheskaya st., 26, St. Petersburg, Russia}
\author{C. Cancellieri}
 \email{a.v.druzhinin@mail.ioffe.ru}
\affiliation{Empa, Swiss Federal Laboratories for Materials Science and Technology, \"{U}berlandstrasse 129, D\"{u}bendorf, 8600, Zurich, Switzerland}

\date{\today}

\begin{abstract}
A model for predicting the residual stress gradient in a thin film segment is developed on the basis of the theory of dislocation pile-ups. The initial shear stress within the film is relaxed via the formation of a pile-up of screw dislocations against the impenetrable film-substrate interface. Plastic strain is related to the dislocation density, leading to a fundamental equation, which links the residual stress to this density. The distribution of dislocations within the pile-up for an arbitrary, non-uniform residual stress profile is derived analytically by applying the force balance condition. This results in a singular integro-differential equation for the residual stress profile. The equation is solved numerically by a collocation method for various initial stress distributions: constant, linear, parabolic, and exponential functions. The solutions demonstrate that the established residual stress profile strongly depends on the film segment's thickness-to-width ratio and the initial stress distribution. As this ratio increases, stress relaxation becomes more effective away from the film-substrate interface. In all cases, equilibrium requires a pile-up containing dislocations with both positive and negative Burgers vectors. The total number of dislocations and their density distribution vary significantly with the initial stress profile. This model provides a critical step towards more complex models of residual stress formation in constrained material systems, specifically, thin films.
\end{abstract}

\maketitle

\section{Introduction}

Residual stresses are typical in polycrystalline thin films due to far from equilibrium conditions of deposition \cite{Abadias2018}. Evolution towards an equilibrium, namely stress relaxation, is harmful to the as-deposited film microstructure. It can affect the structural integrity of the substrate-coating system, leading to film cracking \cite{LI2007}, buckling \cite{FAOU2017}, blistering \cite{MALERBA2016} and delamination \cite{ZHANG2020}. The degradation of the film microstructure is crucial for industrial applications because it limits the performance and durability of the manufactured material. Residual stresses also have a pronounced impact on the degradation of multilayer films with nano-sized layers. Unique for multilayers is the outflow of one of the components from the bulk to the surface of a multilayer, which is a variation of diffusion creep in layers. This phenomenon was reported in multilayers of different material systems, namely Cu/W \cite{DRUZHININ2019, MOSZNER2016}, Ag/AlN \cite{DRUZHININ2023, Chiodi2016}, Cu/Nb \cite{YEOM2025}, Ag-Cu/AlN  \cite{Araullo-Peters2019}. Therefore, controlling the initial stress state and the relaxation process is crucial for predicting the stability of the film's microstructure.

The mechanisms of stress generation depend on the stage of film growth. There is a typically observed tensile-compressive stress transition with an increase in the thickness of the film \cite{Abadias2018}. In the initial stage of growth, when only distinct nuclei are formed, stress can be slightly compressive due to the effect of surface stress \cite{ABERMANN1990}. Upon growth, when coalescence is initiated, as was proposed by Hoffman \cite{HOFFMAN1976}, the generation of tensile stress is due to the decrease in the surface energy of the impinging grains by forming the grain boundary. The origin of compressive stress is believed to be due to the diffusion of atoms into the grain boundaries to decrease the chemical potential in relation to the adatom state. Based on this approach, Chason in his work Ref. \cite{Chason2012} derived the kinetic model, which nicely described the shape of the experimentally observed curves of the average stress evolution. Recently, in Ref. \cite{SUChason2025}, Chason and coauthors improved the model, which fits the experimental data quite well. 

However, most experiments and models are focused on the investigation of the average stress in the film, but not on the distribution (gradient) of stress. However, residual stress gradients are not less important. For instance, residual stress distribution defines the nature of creep in multilayers, and it can act as a means of adjusting the surface outflow during heat treatment \cite{DRUZHININ2021}. A number of experimental methods have been developed to analyze residual stress gradients in thin films, e.g., standard out-of-plane XRD \cite{Klaus2013,Fischer2014}, in-plane GI-XRD \cite{Cancellieri2021a},  FIB-DIC \cite{KORSUNSKY2018}, synchrotron nano-diffraction \cite{KECKES2018}. Despite this, there is a lack of analytical models to predict the profile of residual stress in a film.

The origin of residual stress gradients is believed to be closely linked to the constraint of film plasticity. The film-substrate interface plays a key role in limiting plasticity in a film by establishing the barrier for gliding dislocations. Generally, additional force is required to overcome the stress barrier at the interface between two materials (e.g., see Refs. \cite{WANG2012,HOAGLAND2004,WANG2008,WANG2011}). This would result in a dislocation pile-up in front of the substrate and a non-uniform stress relaxation in a film volume. The latter is behind the origin of the residual stress profile after a sufficient amount of time has passed after deposition. Thus, relying on a relevant mathematical description of dislocation pile-ups near the interfaces, the residual stress profile can be modelled.

The type of gliding dislocation (screw or edge) that relaxes the stress depends on the orientation of the dislocation line and the shear strain in a slip plane. To relax the initial strain, the Burgers vector should be collinear with the displacement vector, which defines the component of the strain tensor in a slip plane. The initial elastic strain can be relieved by screw or edge dislocation, but with different dislocation lines. Considering the problem of residual stress in a film, dislocations must glide towards the barrier, so the model has to be formulated with the relevant type of dislocation. From a mathematical point of view, working with a screw dislocation pile-up is easier, as the stress fields are simpler functions. Relaxation by a mixed dislocation requires taking into account the climb, which also complicates the mathematical description of the self-stress in a pile-up.

Dislocation pile-ups near the interfaces have been widely investigated within the model of continuously distributed dislocations. The distribution of screw dislocations in a pile-up near a bimetallic interface was solved analytically by Barnett \cite{BARNETT1967} and Kuang and Mura \cite{Kuang1968}. Kuang and Mura \cite{Kuang1968} also got a solution for edge dislocation pile-up. Afterwards, Kuang and Mura \cite{Kuang1969} analytically solved the problem of pile-ups near a free surface. Tucker \cite{Tucker1973} and Smith \cite{SMITH1973} applied the continuously distributed dislocation model for analysis of a screw dislocation pile-up near an inclined surface in two-phase materials. In Refs. \cite{LIU2014, AKARAPU2013}, the authors solved the problem of a double-ended edge dislocation pile-up for non-uniform stress profiles. Chang and Mura \cite{CHANG1987} got the solution to the problem of pile-up of dislocations formed by the emission of dislocations from a crack tip. Thereby, there is a rich background in a theory of dislocation pile-ups that can be applied to solve the problem of residual stress profile.

Despite extensive work on stress evolution and relaxation, to the best of our knowledge, there are no published analytical models that predict spatial (depth-resolved) residual stress profiles in a thin film based on dislocation-driven plasticity. The principal goal of this paper is to model the residual stress profile in a thin film segment within the dislocation pile-up theory for the first time. By establishing the model, we determined the equation, relating the residual stress and the plastic strain. We developed the expression for a dislocation density distribution in a pile-up for an arbitrary stress profile. Combining these results, an integro-differential equation was derived, which solution is the spatial distribution of residual stress. We solved this equation numerically for different distributions of initial stress in a film segment. The features of the solutions as well as the limitations of the model are discussed.

\section{Formulation of the model}

\begin{figure*}[ht]
    \centering
    \includegraphics[width=0.8\textwidth]{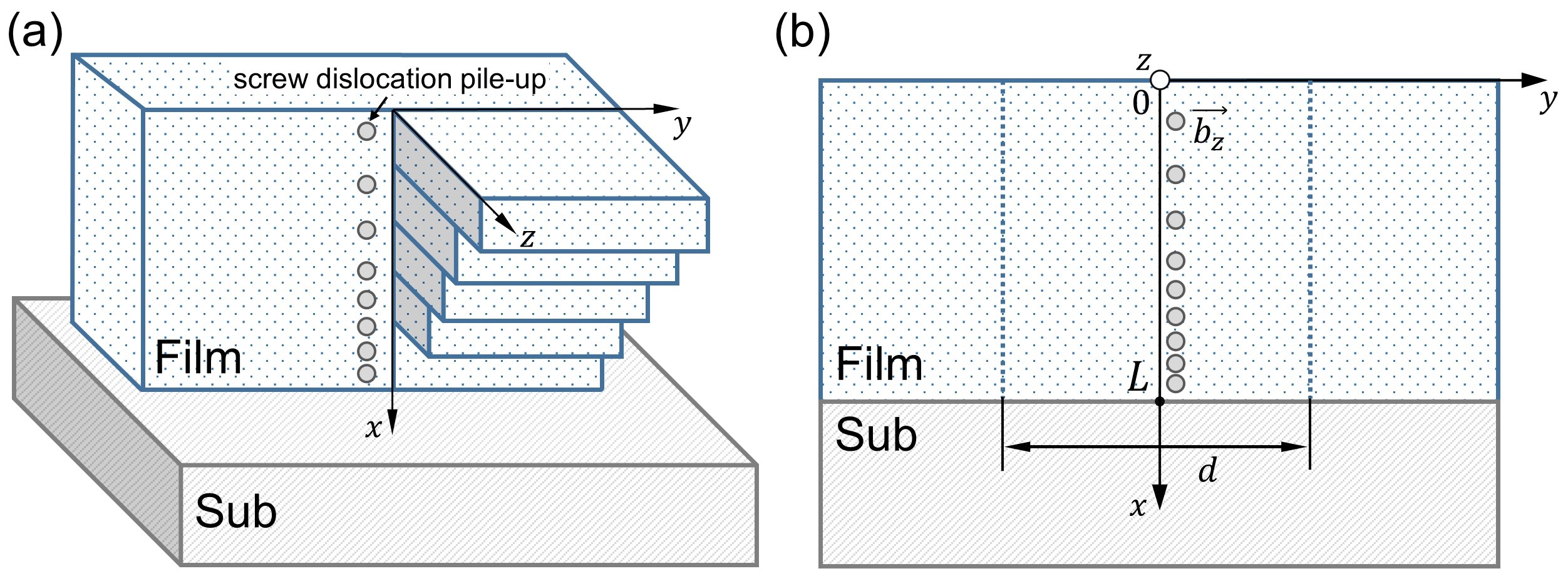}
    \caption{Schematics of the dislocation pile-up model. (a) Visual scheme of a distribution of plastic strain when the pile-up has formed. (b) Geometric parameters used in the model.}
    \label{fig:scheme}
\end{figure*}

Let us consider the following simple one-dimensional model. The segment of a thin film of width $d$ is placed on the substrate surface with the slip plane perpendicular to the film surface. In the initial state, the film is sheared with $\sigma^{0}_{zy}(x)$, which does not depend on the $y$ coordinate. Thus, the residual strain is a function of $x$ only, i.e., $\varepsilon_{zy}(x)$. To relax the initial stress, screw dislocations are formed. The Burgers vectors of dislocations are constant and equal to $b_z$. The substrate acts as an impenetrable obstacle, so the dislocation pile-up is formed. Since
the plasticity near the film-substrate interface is limited,
the initial stress is non-uniformly relaxed, and the residual stress profile is established. The residual stress profile depends on the distribution of dislocations in the pile-up. The distribution of dislocations in the pile-up is a continuous function with the dislocation density per unit length, $n(x)$. 

Initially, all the strain in the film is elastic, $\varepsilon_{zy}^{0}$. Plastic strain increments due to the slip of dislocations. The residual elastic strain is the difference between the initial elastic strain and the plastic strain. Plastic strain at point $x$ on the slip plane, $\varepsilon_{zy}^{pl}(x)$, depends on the amount of dislocations, $N(x)$, which have slipped through this point. Each dislocation contributes to the total plastic strain by the magnitude of its Burgers vector $b_z$. The total plastic strain at $x$ is $\varepsilon^{pl}_{zy}(x)=\frac{b_z\,N(x)}{d}$. 

Therefore, the residual strain in the film segment: 

\begin{equation}
\varepsilon^{res}_{zy}(x)=\varepsilon^{0}_{zy}(x)-\frac{b_z\,N(x)}{d}
    \label{eq1}
\end{equation}

Elastic strains can be expressed through the stresses by using Hooke's law:

\begin{equation}
\frac{d}{2\mu} \sigma^{res}_{zy}(x)=\frac{d}{2\mu} \sigma^{0}_{zy}(x)-b_z\,N(x)
    \label{eq2}
\end{equation}
Here $\mu$ is the shear modulus of a thin film segment.

The number of dislocations that have passed the point $x$, can be derived using the dislocation density, $n(x)$: 

\begin{equation}
N(x)=\int_{x}^{L} n(x')\,dx',
    \label{eq3}
\end{equation}
where $n(x')=\pm\frac{1}{b_z}\frac{db_z}{dx'}$.

Expression for dislocation density, including the Burgers vector:

\begin{equation}
b_z\,N(x)=\int_{x}^{L} b(x')\,dx',
    \label{eq4}
\end{equation}
with $b(x')=b_z\,n(x')$.

Taking the spatial derivative of both sides of Eq. (\ref{eq2}) and considering $\frac{d[b_z\,N(x)]}{dx}=-b(x)$, the following equation in relation to the dislocation density can be derived:

\begin{equation}
\frac{d\sigma^{res}_{zy}(x)}{dx} \frac{d}{2\mu}=\frac{d\sigma^{0}_{zy}(x)}{dx} \frac{d}{2\mu}+b(x)
    \label{eq5}
\end{equation}

In the dimensionless case:

\begin{equation}
\frac{d\sigma_{zy}^{res}(\eta)}{d\eta} \frac{d}{2\mu}=\frac{d\sigma_{zy}^{0}(\eta)}{d\eta} \frac{d}{2\mu}+L\,b(\eta),
    \label{eq6}
\end{equation}
where $\eta=\frac{x}{L}$.

When residual stress is established, the system reaches an equilibrium state, which requires the balance of forces acting on individual dislocations in a pile-up. This requirement is met when the external force is equal to the forces imposed by other dislocations, i.e., due to the self-stress of the dislocations. In the present model, we can consider the residual stress as an external force imposed by a film volume surrounding the considered segment. For example, it could arise from the atoms incorporated in grain boundaries upon growth, which is typical for polycrystalline films (see Ref. \cite{Chason2012}). Thus, dislocation density is also a function of $\sigma^{res}_{zy}(x')$, i.e., $b(\sigma^{res}_{zy}(x'),x)$ . Knowing the functional relationship for $b(\sigma^{res}_{zy}(x'),x)$, the equation for $\sigma^{res}_{zy}(x)$ is determined for the specified distribution of initial stress, $\sigma^{0}_{zy}(x)$.

\section{Derivation of $n(\sigma^{res}_{zy}(x'),x)$}

The main difficulty in solving Eq. (\ref{eq5}) is the derivation of the proper expression for the dislocation density. As mentioned above, the expression for $n(\sigma_{zy}^{res}(x'),x)$  is determined by the force balance. The force imposed on an individual dislocation at point $x$ in the pile-up is derived by integrating the self-stress of other dislocations \cite{hirth1992theory}. For the case of a thin film on a substrate, an expression for a self-stress of a dislocation has to consider the presence of the free surface and the difference of elastic moduli of the film and substrate. The modified expression is determined by introducing image dislocations. The general expression for the self-stress of a screw dislocation in a film on a substrate was derived by Chou \cite{Chou1966} (here "film" is identical to the term "lamella inclusion" in the original paper). However, it is not an easy task to use this solution, since it contains an infinite sum of fractional terms with spatial coordinates, making Eq. (\ref{eq5}) quite complex. A more feasible task is to use the self-stress of a screw dislocation near the free surface, i.e., assuming that the difference of the film and the substrate shear moduli is negligible. In Ref. \cite{Kuang1969}, Kuang and Mura solved the problem of (screw and edge) dislocation pile-up near the free surface. In this case, self-stress of a dislocation includes the fractional term related to the image dislocation of an opposite sign placed above the free surface. In the present work, we use the approach of Kuang and Mura to derive the expression for the dislocation density, that is, the difference in shear moduli of the film and the substrate is assumed to be small. Following their derivation, an integral expression for $n(\sigma^{res}_{zy}(x'),x)$ (or $b(\sigma^{res}_{zy}(x'),x)$) can be derived, so Eq. (\ref{eq5}) transforms to an integro-differential equation. Thus, the solution of Eq. (\ref{eq5}) begins with the definition of dislocation density as a function of residual stress, which is described in this section.

The balance of forces acting on a screw dislocation is

\begin{equation}
\sigma_{zy}^{res}(x)+\frac{\mu\,b_z}{2\pi}\int^{L}_{0} \left[\frac{1}{x-\xi}-\frac{1}{x+\xi} \right]\,n(\xi)\,d\xi=0,
    \label{eq7}
\end{equation}
where $\xi$ is the position of screw dislocations in the pile-up.

To introduce the dimensionless form of Eq. (\ref{eq7}), we will use a parameterization similar to that used by Kuang and Mura in their work \cite{Kuang1969}. Eq. (\ref{eq7}) can be written as

\begin{equation}
\sigma_{zy}^{res}(t)+\frac{\mu\,b_z}{2\pi}\int^{1}_{0} \frac{1}{\eta}\left[\frac{1}{t/\eta-1}-\frac{1}{t/\eta+1} \right]\,n(\eta)\,d\eta=0,
    \label{eq8}
\end{equation}
where $t=\frac{x}{L}$ and $\eta=\frac{\xi}{L}$.

Moving $b_z$ inside the integral and assigning $b(\eta)=b_z\,n(\eta)$ and $\gamma=\mu/2\pi$, we get exactly Eq. 2 of the Kuang and Mura paper \cite{Kuang1969}:

\begin{equation}
\sigma_{zy}^{res}(t)+\gamma\int^{1}_{0} \frac{1}{\eta}\left[\frac{1}{t/\eta-1}-\frac{1}{t/\eta+1} \right]\,b(\eta)\,d\eta=0,
    \label{eq9}
\end{equation}

We follow the solution steps of the original papers of Kuang and Mura \cite{Kuang1969,Kuang1968}. The authors solved the integral equation Eq. (\ref{eq9}) by the Wiener-Hopf technique after the Mellin transform. The final solution obtained in the work \cite{Kuang1969} is a distribution of dislocations $b(\eta)$ in relation to the applied constant stress. However, in our case, the main task is to find $b(\eta)$ for the general case of an applied stress of an arbitrary spatial distribution, $b(\sigma^{res}_{zy}(t),\eta)$. The notation used is identical to that in Refs. \cite{Kuang1968,Kuang1969}.

The dislocation distribution function is derived by means of the inverse Mellin transform:
\begin{equation}
b(\eta)=\frac{1}{2\pi i} \int_{\tau-i\infty}^{\tau+i\infty} B_{-} (s)\, \eta^{-s}\, ds,
    \label{eq10}
\end{equation}
where $0<\tau<1$ for the strip in the complex $s$ plane (the image part is $-\infty<p<\infty$).

The parameter $B_{-}(s)$ is determined in terms of the variables $C_{-}(s)$ and $K_{-}(s)$ introduced in Refs. \cite{Kuang1969,Kuang1968} as follows

\begin{equation}
    B_{-}(s)=\frac{C_{-}(s)}{K_{-}(s)}
    \label{eq11}
\end{equation}

\begin{widetext}
\begin{equation}
C_{-}(s)=-\frac{1}{2 \pi i} \, \int_{0}^{1} \frac{\sigma^{res}_{zy}(t)}{t}\,dt \int_{c_{1}-i\,\infty}^{c_{1}+i\,\infty} \frac{t^{\lambda} \Gamma(0.5-0.5\lambda)^2}{\pi \sqrt{0.5 \gamma}\, 2^{\lambda+2}\, \Gamma(1-\lambda) (\lambda-s)} \, d\lambda
    \label{eq12}
\end{equation}
\end{widetext}

\begin{equation}
K_{-}(s)=\frac{\pi \sqrt{0.5\,\gamma}\,\Gamma(s)}{2^s \Gamma\left[0.5(s+1) \right]^2}
    \label{eq13}
\end{equation}
Here $c_1$ is enclosed in the in the $(0,1)$ interval.

Kuang and Mura in their original papers first make an integration for $t$, taking $\sigma_{zy}^{res}(t)$ a constant. In contrast, we first integrate the inner integral in Eq. (\ref{eq12}) to maintain an explicit dependence of $C_{-}(s)$ on $\sigma_{zy}^{res}(t)$.

The inner integral in Eq. (\ref{eq12}) can be solved by the residue theorem.  The poles of the integrand are $s$ and 1, 3, 5, 7, ... . Residue at $s$ is:
\begin{equation}
\text{res}(C_{-} (s),s)=\frac{2^{-2 - s} \, t^s \, \Gamma\!\left(\frac{1}{2} - \frac{s}{2}\right)^{\!2}}{\pi \, \sqrt{ 0.5 \gamma} \, \Gamma(1 - s)}
    \label{eq14}
\end{equation}
The sum of residues at poles 1, 3, 5, 7, ... :

\begin{equation}
\sum_{n=1}^\infty \text{res}(C_{-} (s),n)=\sum_{n=1}^\infty \frac{(2n-2)! \, t^{2n-1}}{\pi \sqrt{0.5 \gamma} \, [(n-1)!]^2\,2^{2n-1} (s - (2n - 1))}
    \label{eq15}
\end{equation}

\begin{widetext}
\begin{equation}
C_{-} (s)=\frac{1}{\pi \sqrt{0.5 \gamma}}\int_0^1 \frac{\sigma^{res}_{zy}(t)}{t} \left(
  \frac{2^{-2-s} t^s \Gamma\left(\frac{1}{2} - \frac{s}{2}\right)^2}{\Gamma(1-s)}
  + \sum_{n=1}^\infty \frac{(2n-2)! \, t^{2n-1}}{[(n-1)!]^2\,2^{2n-1} (s - (2n - 1))}
\right) dt
    \label{eq16}
\end{equation}
\end{widetext}

For a constant $\sigma^{res}_{zy}(t)$ it reduces to $\frac{\sigma^{res}_{zy}}{4s\sqrt{0.5 \gamma}}$, i.e., the expression obtained in Ref. \cite{Kuang1969}. Returning to Eq. (\ref{eq10}):

\begin{widetext}
\begin{multline}
   b(\eta)=\frac{1}{2\pi i} \int_0^1 \frac{\sigma^{res}_{zy}(t)}{t}\left( \frac{1}{\pi \sqrt{0.5 \gamma}} \int_{\tau-i\infty}^{\tau+i\infty}  \left(
  \frac{2^{-2-s} t^s \Gamma\left(\frac{1}{2} - \frac{s}{2}\right)^2}{\Gamma(1-s)}
  + \sum_{n=1}^\infty \frac{(2n-2)! \, t^{2n-1}}{[(n-1)!]^2\,2^{2n-1} (s - (2n - 1))} \right)  \frac{2^{s}\, \Gamma(\frac{1}{2}(s+1))^{2}}{\pi \sqrt{0.5 \gamma}\,\Gamma(s)} \eta^{-s}\, ds \right) dt
      \label{eq17} 
\end{multline}
\end{widetext}

The inverse Mellin transform of the integrand function by the residue theorem gives:

\begin{widetext}
\begin{equation}
    b(\eta)=\int_{0}^{1} \sigma^{res}_{zy}(t) \left( \frac{1}{t} \left( \frac{\eta t}{(\eta^2 - t^2)\pi^2 \, 0.5 \, \gamma} + \sum_{n=1}^{\infty} \sum_{k=1}^{\infty} \frac{(2n-2)! t^{2n-1}}{((n-1)!)^2 \, \pi^2 \,  0.5 \,\gamma \,2^{2n-1}} \frac{(2k-1)!! \eta^{2k-1}}{(2k-2)!! (k-1+n)} \right) \right) dt
    \label{eq18}
\end{equation}
\end{widetext}

Determining the closed form of the double sum, we get the final expression for the dislocation distribution function:

\begin{widetext}
\begin{equation}
    b\left(\sigma^{res}_{zy}(t),\eta\right)=\int_{0}^{1} \sigma_{zy}^{res}(t) \left( -\frac{2\eta}{\gamma \pi^{2} (t^{2} - \eta^{2})} \sqrt{\frac{t^{2} - 1}{\eta^{2} - 1}}\right) dt
    \label{eq19}
\end{equation}       
\end{widetext}

The total amount of dislocations is

\begin{equation}
    N=\frac{L}{b_z}\,\int_{0}^{1} b(\eta) \, d\eta
    \label{eq20}
\end{equation}

\section{Numerical solution of the integro-differential equation}

Using the expression determined for $b(\eta)$, Eq. (\ref{eq6}) is

\begin{widetext}
\begin{equation}
\frac{d\sigma_{zy}^{res}(\eta)}{d\eta} \frac{d}{2\mu}=\frac{d\sigma_{zy}^{0}(\eta)}{d\eta} \frac{d}{2\mu}+ L \int_{0}^{1} \sigma_{zy}^{res}(t) \left( -\frac{2\eta}{\gamma \pi^{2} (t^{2} - \eta^{2})} \sqrt{\frac{t^{2} - 1}{\eta^{2} - 1}}\right) dt,
    \label{eq21}
\end{equation}    
\end{widetext}

where $\gamma=\mu/2\pi$.

After the rearrangement of coefficients:

\begin{widetext}
\begin{equation}
\frac{d\sigma_{zy}^{res}(\eta)}{d\eta}=\frac{d\sigma_{zy}^{0}(\eta)}{d\eta}+ k \int_{0}^{1} \sigma_{zy}^{res}(t) \left( -\frac{2\eta}{\gamma_1 \pi^{2} (t^{2} - \eta^{2})} \sqrt{\frac{t^{2} - 1}{\eta^{2} - 1}}\right) dt,
    \label{eq22}
\end{equation}    
\end{widetext}

where $k=2L/d$ and $\gamma_1=1/2\pi$. 

Thus, the final equation is a singular integro-differential equation with the following kernel:

\begin{equation}
K(\eta,t)=-\frac{2\eta}{\gamma_1 \pi^{2} (t^{2} - \eta^{2})} \sqrt{\frac{t^{2} - 1}{\eta^{2} - 1}}
    \label{eq23}
\end{equation}

It should be noted that this equation does not depend on the magnitude of the film shear modulus $\mu$. The dimensionless parameter $k$ measures the ratio between the thickness and the width of the film segment. As mentioned in the next section, the magnitude of $k$ defines the effectiveness of stress relaxation in a film volume. The parameters used in the model are summarized in Table \ref{tab:table1}.   

\begin{table}[h!]
\caption{\label{tab:table1}Parameters used in the model}
\begin{ruledtabular}
\begin{tabular}{cc}
Parameter&Definition\\
\hline
$L$ & film thickness\\
$d$ & film segment width\\
$\mu$ & film shear modulus\\
$\gamma$ & $\mu/2\pi$\\
$\gamma_1$ & $1/2\pi$\\
$k$ & $2L/d$\\
\end{tabular}
\end{ruledtabular}
\end{table}

The boundary condition is the following. At the film-substrate interface $\eta=1$, there is a barrier stress to transmit the dislocation to the substrate. We assume that the initial stress in a film segment is less than the barrier stress. Thus, as the substrate-film interface is approached, the amount of dislocations at each point $\eta$, slipping towards the substrate, decreases. In the limit $\eta \rightarrow 1$, the stress is completely unrelaxed, corresponding to the value of the initial stress at this point, i.e., $\sigma(1)=\sigma_0$.

Unfortunately, we failed to solve Eq. (\ref{eq22}) analytically even for the constantly distributed initial stress, i.e., $\sigma_{zy}^{0}(\eta)/d\eta=0$. To derive the main features of the residual stress profiles and pile-ups, we solved this equation numerically by the collocation method. This method approximates a function by a linear combination of basis functions and ensures that the differential equation is exactly satisfied at a finite set of selected points. In our case, we used polynomial functions of the ninth order. Solving numerically the integro-differential equation, we get the coefficients for the polynomial function used. The values of tensile initial stress and shear moduli are taken as an example to illustrate the trends. The results for compressive initial stresses are physically identical but mirrored relative to the $\eta$ axis.

It is hard to say unambiguously what type of initial stress distribution should be taken to simulate the state of a grown film correctly. Intuition suggests several limiting cases with certain initial stress distributions. When the grains are large enough, the weight of grain boundaries is minor, so the coherency stress due to lattice misfit and interface stress would govern the state of residual stress in a film. This could correspond to the constant initial stress. In contrast, when grains are small, which is more relevant to sputtering methods, grain boundaries are expected to have an effect on the residual stress within a grain, leading to a non-uniform initial stress profile.  Its shape must depend on the interplay of atomic mobility and the functional dependence of the chemical potential on the position in the grain boundary \cite{Chason2012}. The high mobility of the atoms could smooth the concentration gradient in a grain boundary, tending to a constant distribution of initial stress in the as-deposited state. If the excess of chemical potential at the surface of the growing film changes upon deposition and the atomic mobility is low, we expect non-uniform distributions of the initial stress.

\begin{figure*}[ht]
    \centering
    \includegraphics[width=0.9\textwidth]{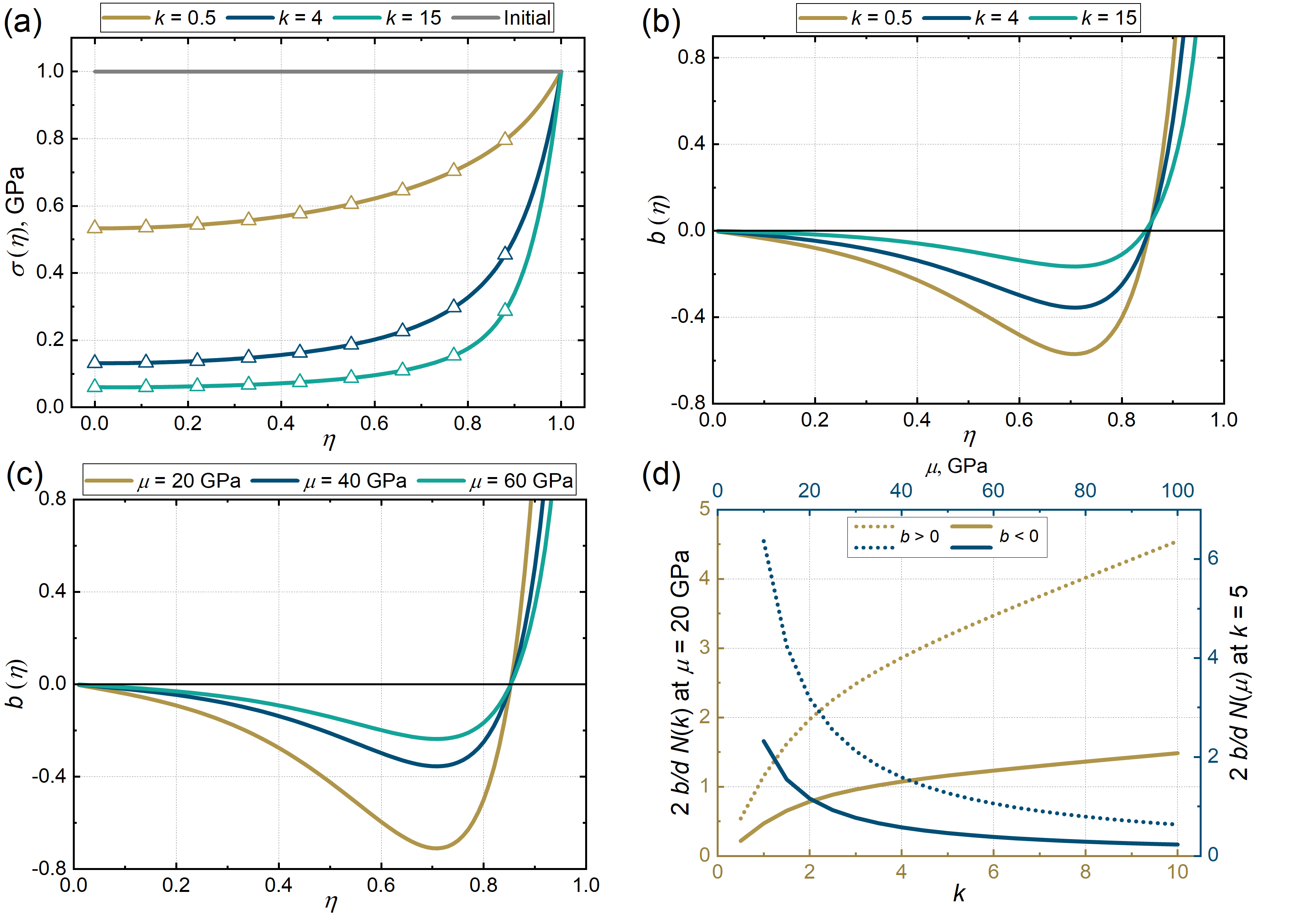}
    \caption{Numerical solution for the case of a constant initial stress. (a) Residual stress distribution for different $k=2L/d$ values. (b, c) Dislocation density distribution for different $k$ and $\mu$ values. (d) Total amount of dislocations, $N$, (multiplied by $\frac{2b_z}{d}$) as a function of $\mu$ and $k$. Collocation points are highlighted by triangles. The grey curve is the profile of initial stress}
    \label{fig:const}
\end{figure*}

Thus, to demonstrate the features of residual stress profiles, we have selected four functions to describe the spatial distribution of initial stress, namely constant, linear, parabolic and exponential. The curvatures of exponential and parabolic profiles are the opposite. We also derived solutions for the initial stress profile, which changes the sign in the film volume. The features of the derived residual stress profiles are discussed in the Discussion section.

\subsection{Constant initial stress}

The results for a constant initial stress are presented in Fig. \ref{fig:const}. In Fig. \ref{fig:const}a, residual stress curves for different magnitudes of $k$ are shown. As expected intuitively, with an increase in the thickness-to-width ratio, $k=2L/d$, there is a stronger stress relaxation in the film volume away from the substrate surface, so the residual stress curves are close to zero stress. In Fig. \ref{fig:const}b, the dislocation density distributions for the same values of $k$ are shown. It is significant to note that dislocations with opposite signs of the Burgers vector are formed. Dislocations with $b_z >0$ are concentrated at the film-substrate interface, whereas dislocations with $b_z<0$ are smeared across the film thickness. At approximately $\eta \approx 0.85$, the density of dislocations approaches zero, changing the sign. The peak of dislocation density is at $\eta \approx 0.7$, gradually decreasing toward the film surface. In Fig. \ref{fig:const}c, the dislocation density is shown as a function of the shear moduli of the film, $\mu$. The dislocation density decreases with increasing $\mu$. It is intuitively correct, since the self-stress of a dislocation is linearly proportional to the magnitude of the shear modulus; thereby, a smaller number of dislocations are needed to balance an externally applied stress. In Fig. \ref{fig:const}d, the total number of dislocations (multiplied by $2b_z/d$) is shown as a function of $\mu$ and $k$. More dislocations with $b_z>0$ are formed upon relaxation. The total number of dislocations with both signs of $b_z$ increases with increasing the magnitude of $k$. Notably, the number of dislocations decreases non-linearly with an increase of the $\mu$ magnitude.

\subsection{Linear initial stress}

\begin{figure*}[ht]
    \centering
    \includegraphics[width=0.9\linewidth]{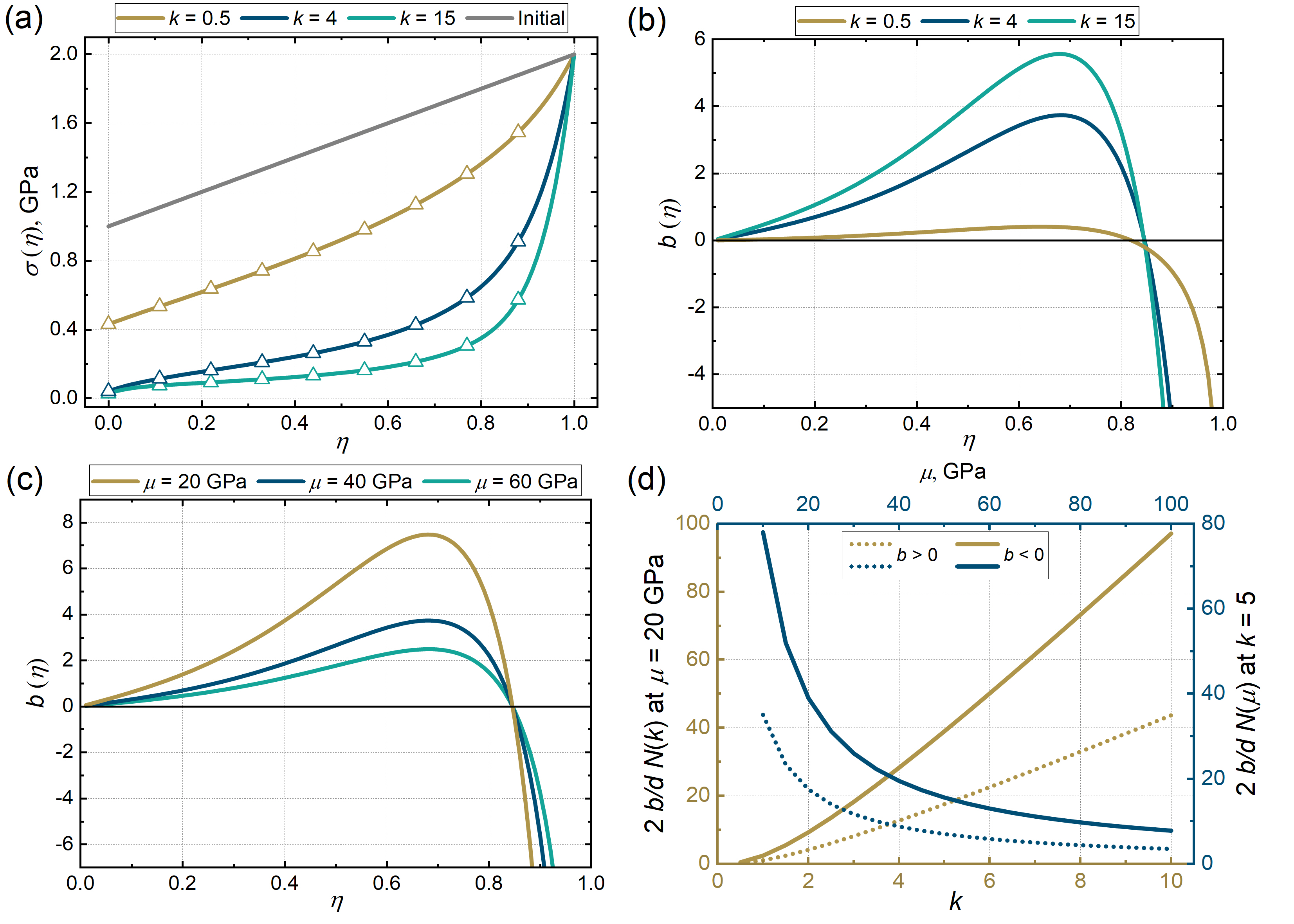}
    \caption{Numerical solution for the case of a linearly distributed initial stress with a constant sign. (a) Residual stress distribution for different $k=2L/d$ values. (b, c) Dislocation density distribution for different $k$ and $\mu$ values. (d) Total amount of dislocations, $N$, (multiplied by $\frac{2b_z}{d}$) as a function of $\mu$ and $k$. Collocation points are highlighted by triangles. The grey curve is the profile of initial stress}
    \label{fig:linear1}
\end{figure*}

The results for a linearly distributed initial stress of a constant sign are presented in Fig. \ref{fig:linear1}. Likewise the case of a constant initial stress, the stress is intensively relaxed by increasing $k$ (see Fig. \ref{fig:linear1}a). Noteworthy, residual stress near the film surface (at $\eta \approx0$), is almost similar for $k$ equals 4 and 15. The dislocation density as a function of $\eta$ is the opposite in relation to the case of constant initial stress. Now, negative dislocations ($b_z<0$) are concentrated near the substrate surface, and the major part of the film volume is filled with positive dislocations ($b_z>0$). There is a similar tendency for the dislocation density to decrease with an increase of $\mu$ (see Fig. \ref{fig:linear1}c). Fig. \ref{fig:linear1}d shows that dislocations with $b_z<0$ are dominant. It should be noted that the number of dislocations is considerably larger than in the case of constant initial stress.

\begin{figure*}[ht]
    \centering
    \includegraphics[width=0.9\linewidth]{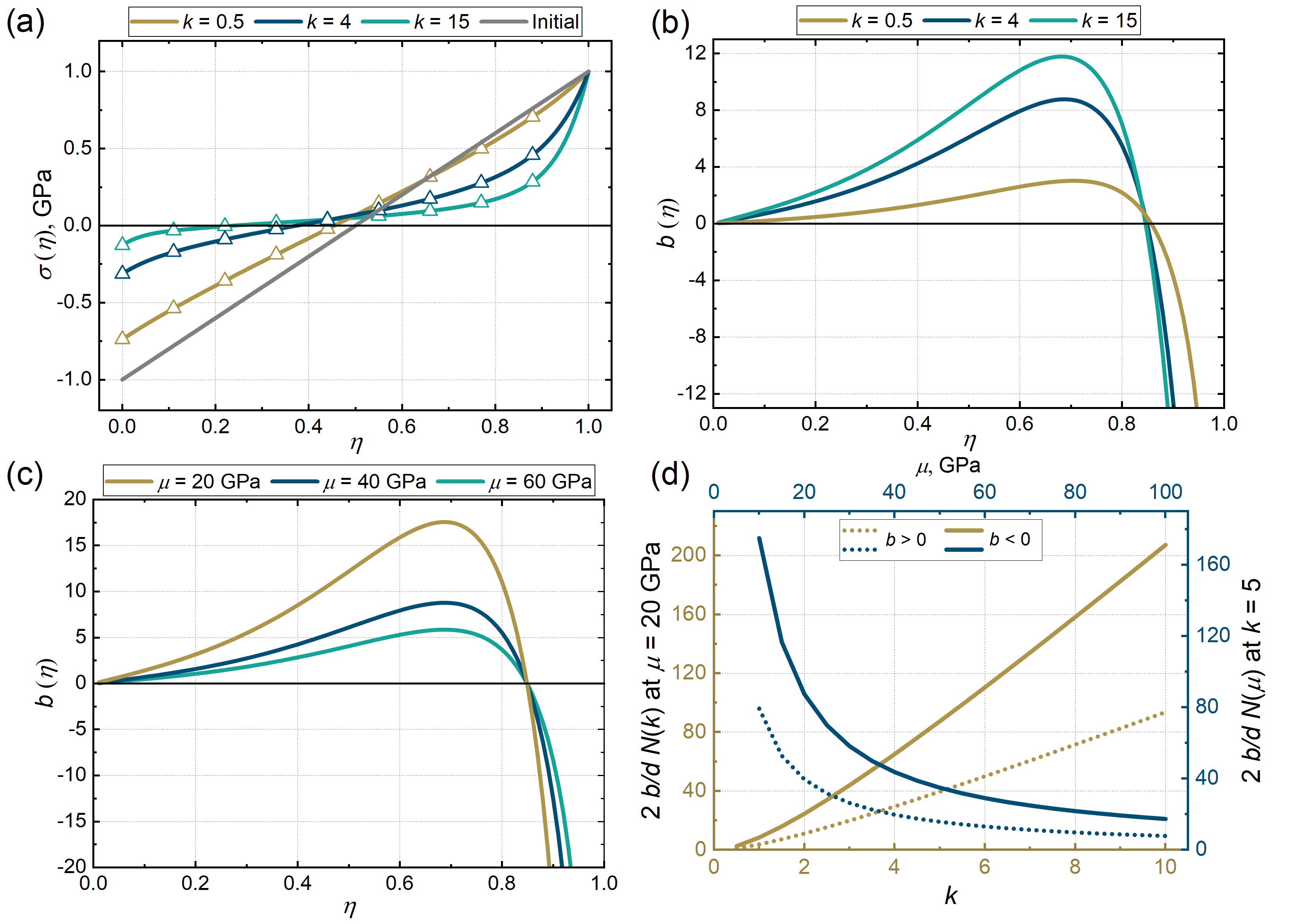}
    \caption{Numerical solution for the case of a linear distribution of initial stress with a changing sign. (a) Residual stress distribution for different $k=2L/d$ values. (b, c) Dislocation density distribution for different $k$ and $\mu$ values. (d) Total amount of dislocations, $N$, (multiplied by $\frac{2b_z}{d}$) as a function of $\mu$ and $k$. Collocation points are highlighted by triangles. The grey curve is the profile of initial stress}
    \label{fig:linear2}
\end{figure*}

Fig. \ref{fig:linear2} shows the results for a linearly distributed initial stress that changes its sign in the volume of the film. In the equilibrium state, there are regions of tensile and compressive stresses (Fig. \ref{fig:linear2}a). The point of zero residual stress is slightly moving to $\eta=0$ as $k$ is increased. The shapes of $b(\eta)$ and $\frac{2b_z}{d}N(k)$ curves are similar to those presented in Fig. \ref{fig:linear1}b-d. However, the magnitudes of the total number of dislocations increase substantially, that is, almost twice.

\subsection{Parabolic initial stress}

\begin{figure*}[ht]
    \centering
    \includegraphics[width=0.9\textwidth]{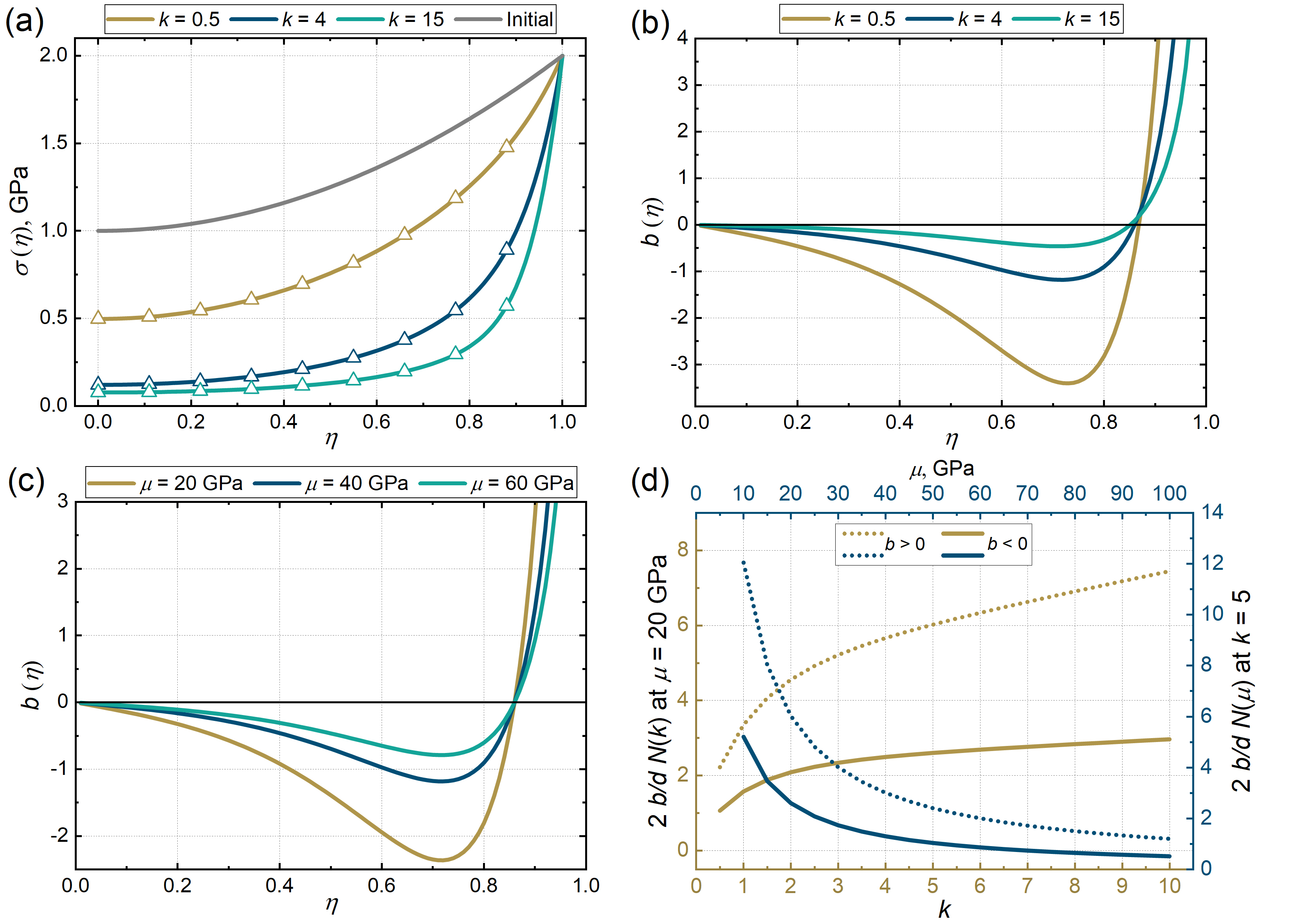}
    \caption{Numerical solution for the case of a parabolic distribution of initial stress with a constant sign. (a) Residual stress distribution for different $k=2L/d$ values. (b, c) Dislocation density distribution for different $k$ and $\mu$ values. (d) Total amount of dislocations, $N$, (multiplied by $\frac{2b_z}{d}$) as a function of $\mu$ and $k$. Collocation points are highlighted by triangles. The grey curve is the profile of initial stress}
    \label{fig:quadratic1}
\end{figure*}

The results of a numerical solution for a parabolic distribution of initial stress with a constant sign are shown in Fig. \ref{fig:quadratic1}. The shape of the residual stress profiles as a function of $k$ corresponds to the linear $\sigma^{0}_{zy}(\eta)$ (Fig. \ref{fig:quadratic1}a). However, now dislocations with $b_z>0$ are major (see Figs. \ref{fig:quadratic1}b and c). At the same time, the total number of dislocations is almost two orders of magnitude smaller, correlated with the results of a constant initial stress. It is notable that the evolution of $\frac{2b_z}{d}N(k)$ at constant $\mu$ has a curvature different from the one observed in the case of a linear initial stress (Fig. \ref{fig:quadratic1}d). 

\begin{figure*}[ht]
    \centering
    \includegraphics[width=0.9\textwidth]{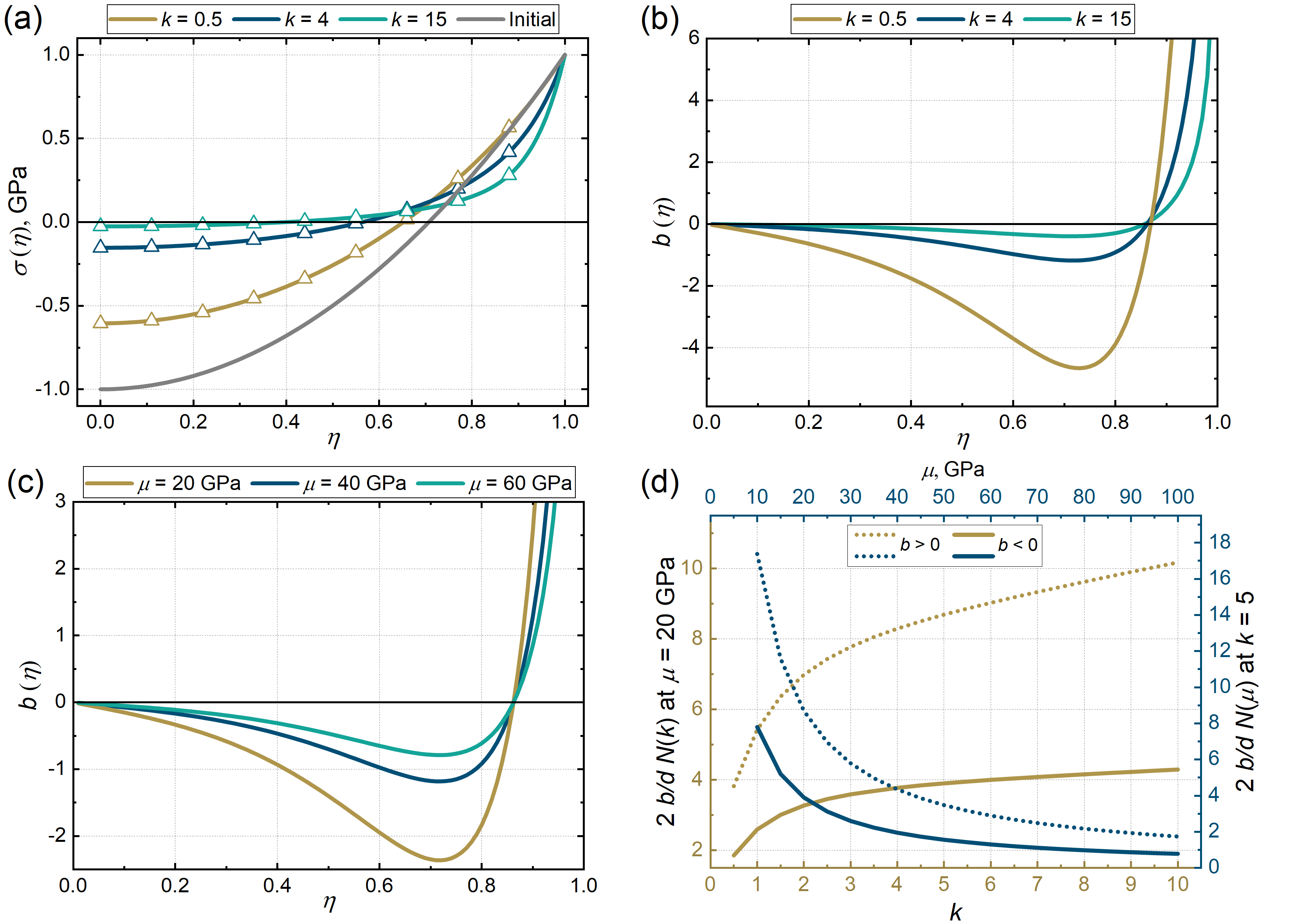}
    \caption{Numerical solution for the case of a parabolic distribution of initial stress with a changing sign. (a) Residual stress distribution for different $k=2L/d$ values. (b, c) Dislocation density distribution for different $k$ and $\mu$ values. (d) Total amount of dislocations, $N$, (multiplied by $\frac{2b_z}{d}$) as a function of $\mu$ and $k$. Collocation points are highlighted by triangles.  The grey curve is the profile of initial stress}
    \label{fig:quadratic2}
\end{figure*}

In Fig. \ref{fig:quadratic2}, the results for the initial parabolic stress with a changing sign are presented. Again, the equilibrium residual stress profiles preserve the compressive and tensile regions, and the point of zero stress again tends to zero as $k$ increases. The nature of the dislocation density distribution $b(\eta)$ as well as the total number of dislocations do not undergo severe changes (Figs. \ref{fig:quadratic2}b-d). 

\subsection{Exponential initial stress}

\begin{figure*}[ht]
    \centering
    \includegraphics[width=0.9\textwidth]{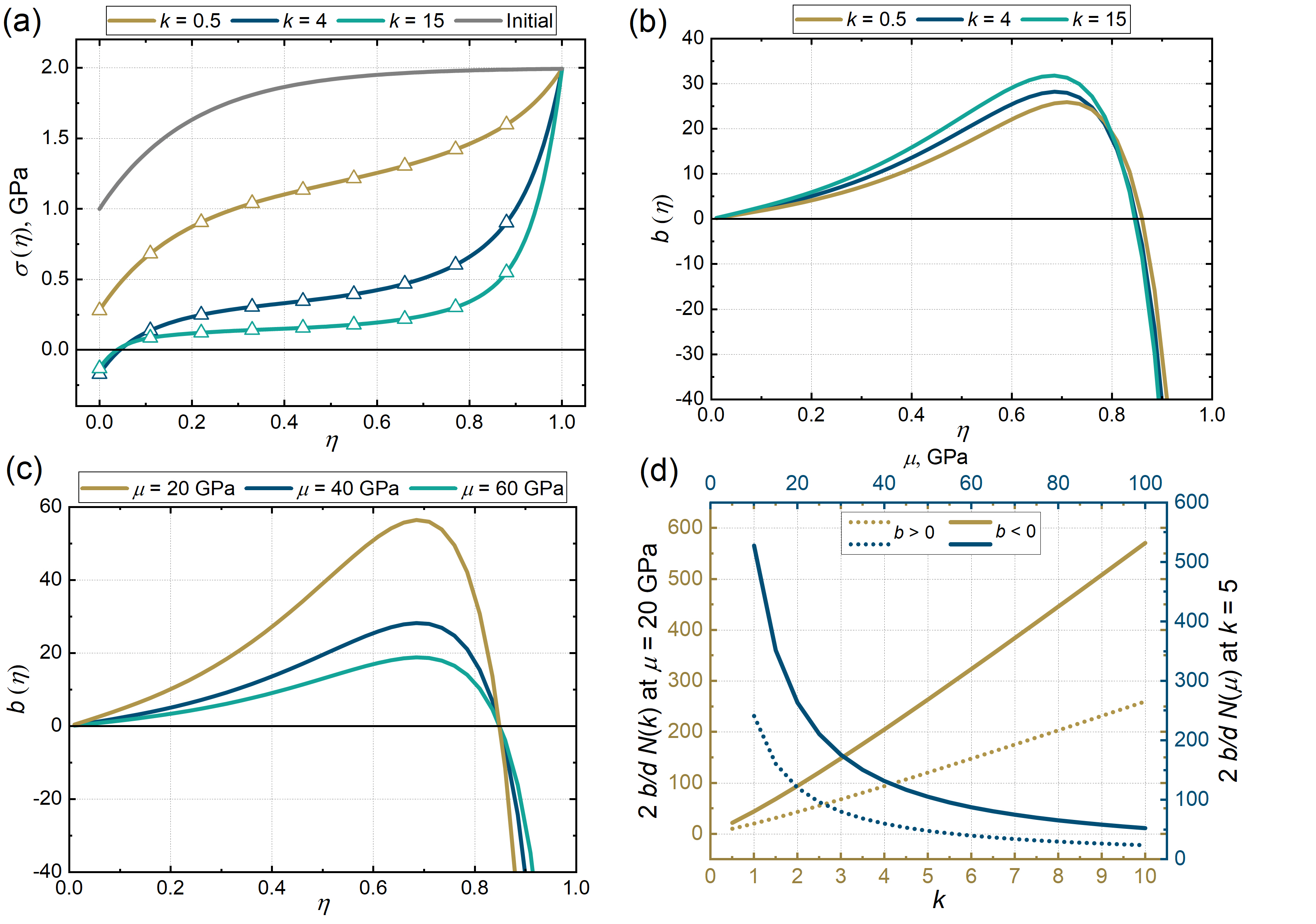}
    \caption{Numerical solution for the case of an exponential distribution of initial stress with a constant sign. (a) Residual stress distribution for different $k=2L/d$ values. (b, c) Dislocation density distribution for different $k$ and $\mu$ values. (d) Total amount of dislocations, $N$, (multiplied by $\frac{2b_z}{d}$) as a function of $\mu$ and $k$. Collocation points are highlighted by triangles. The grey curve is the profile of initial stress}
    \label{fig:exp1}
\end{figure*}

Fig. \ref{fig:exp1} shows the results for the case where the initial stress has an exponential distribution with a constant sign. To reveal the impact of curvature on stress relaxation, an exponential profile with a negative curvature is taken (compared to profiles investigated previously). Analyzing the results in Fig. \ref{fig:exp1}a, it is found that near the surface of the film, the residual stress profiles maintain the curvature of the initial distribution. The curves with $k$ equals 4 and 15 go negative as $\eta \rightarrow 0$. The dislocation distributions consist mainly of dislocations with $b_z>0$ (Figs. \ref{fig:exp1}b,c). The variation of $b(\eta)$ with $k$ is slight and, in fact, is the weakest among the investigated cases. In contrast, the total number of dislocations $\frac{2b_z}{d}N(k)$ is the largest (Fig. \ref{fig:exp1}d). The shape of the $\frac{2b_z}{d}N(k)$ curve as a function of $k$ is similar to the one obtained for a linear distribution of initial stress.

\begin{figure*}[ht]
    \centering
    \includegraphics[width=0.9\textwidth]{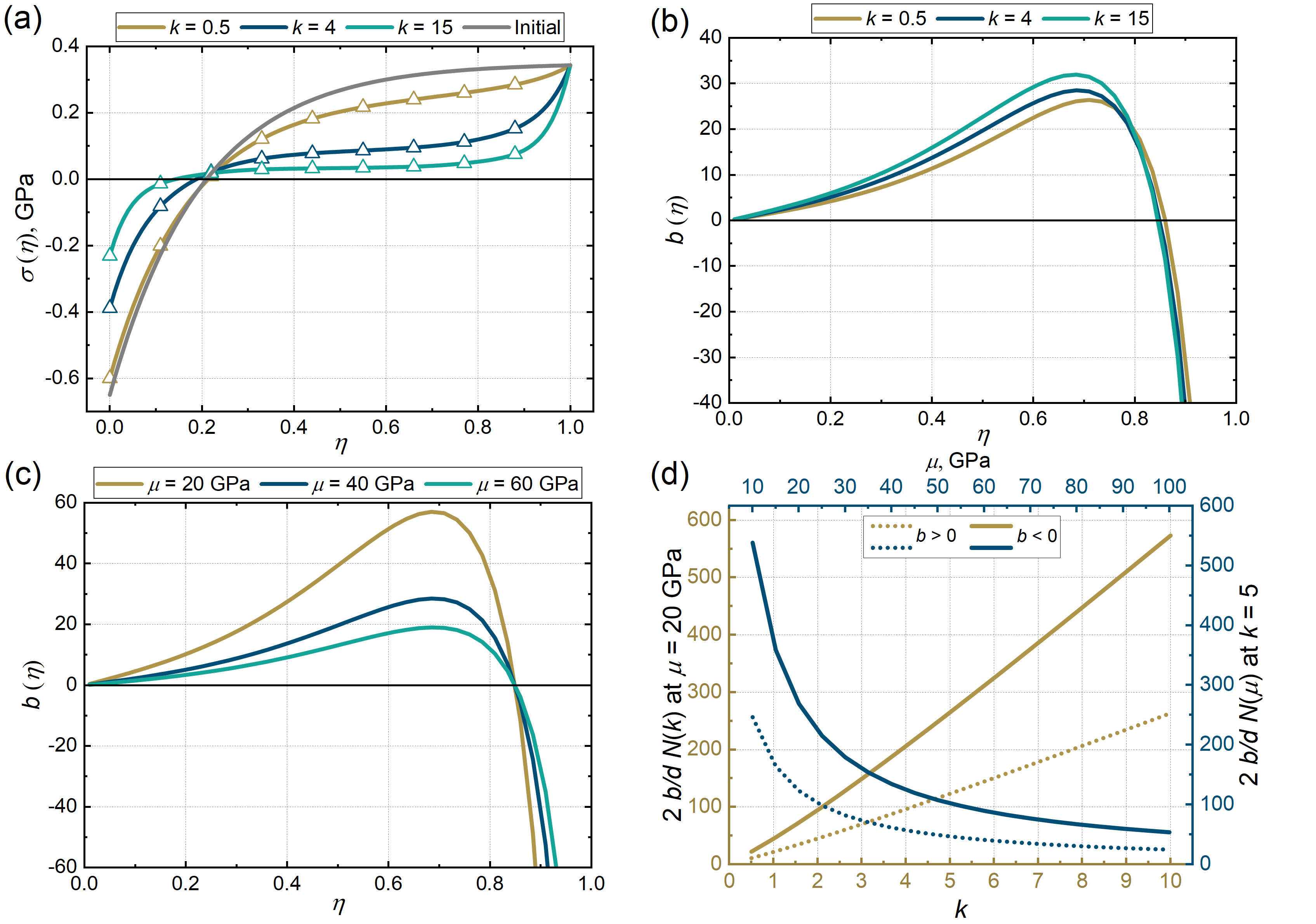}
    \caption{Numerical solution for the case of an exponential distribution of initial stress with a changing sign throughout the film thickness. (a) Residual stress distribution for different $k=2L/d$ values. (b, c) Dislocation density distribution for different $k$ and $\mu$ values. (d) Total amount of dislocations, $N$, (multiplied by $\frac{2b_z}{d}$) as a function of $\mu$ and $k$. Collocation points are highlighted by triangles. The grey curve is the profile of initial stress}
    \label{fig:exp2}
\end{figure*}

The results for the exponential distribution of the initial stress with a changing sign are presented in Fig. \ref{fig:exp2}. The main trends in $b(\eta)$ and $\frac{2b_z}{d}N(k)$ are similar to the described above. As before, the total number of dislocations $\frac{2b_z}{d}N(k)$ is the largest. The residual stress profiles in Fig. \ref{fig:exp2}a show a transition from compressive to tensile stress, reproducing the shape of the initial distribution. However, the point $\eta$, where this transition occurs, is the same for each value of $k$, unlike the cases shown above. 

\section{Discussion}

The necessity of dislocations with Burgers vectors of opposite signs to establish equilibrium residual stress profiles is one of the outcomes of the derived numerical solutions. This is in contrast to classic one-sided pile-ups under a constant applied load, as demonstrated, e.g., in Refs. \cite{hirth1992theory, Kuang1969, Kuang1968}. However, it is not surprising, since the problem considered in this study can be interpreted as plastic deformation in a material with a finite amount of elastic energy. Upon relaxation, this elastic energy is spent to form the pile-up, so a non-uniform external stress distribution evolves in the film. Indeed, in the works cited above, the applied stress remains spatially constant during the pile-up formation process. Therefore, in the equilibrium state, dislocations of only one sign are needed to equilibrate the external force acting on each individual dislocation. In support of the hypothesis mentioned above, similar dislocation pile-ups were found, consisting of dislocations with negative and positive Burgers vectors. These were obtained by solving the problem with applied non-uniformly distributed initial stress profiles in Refs. \cite{AKARAPU2013, LIU2014}.

The point $\eta$, where $b(\eta)=0$, can be regarded as the dislocation source, which emits dislocations of both signs during relaxation of stress. It is worth to note that the dislocation source is always formed near the film-substrate interface, and its location $\eta$ is barely affected by a change in $k$. This location could arise due to the high stress gradients near the film-substrate interface related to the boundary condition. Considering the kinetics, it is expected that several dislocation sources may arise in the rapidly relaxing film. However, since this state is not equilibrium, the largest possible stress relaxation would not be achieved.

Strictly speaking, the boundary condition has to be modified. When the pile-up near the film-substrate interface is dense, the stress in the pile-up head is large enough to overcome the stress barrier for transmitting the dislocation to the substrate. Dislocations cross-slip to the substrate, until the stress near the interface is considerably decreased to become smaller than the barrier stress. Thus, in the equilibrium state, the residual stress at the film-substrate interface must be equal to the barrier stress, which is likely to differ from the initial stress at $\eta=1$. 

For the initial stress distributions examined in this study, the dislocation sign changes only once in the volume of the film. It is likely that solving the equation with the wavy initial stress distribution would result in pile-up with 
$b_z$ changing its sign several times. However, such initial stress distributions are barely expected to be physically real. If the deposition conditions during film growth are kept constant, a monotonous stress profile is intuitively expected, i.e., atoms are sequentially rearranged during deposition, e.g., by penetration to the grain boundaries \cite{Chason2012}, resulting in a force that acts on a grain . Changes in deposition conditions could potentially introduce fluctuations in the initial stress profile. 

It is important to highlight that the residual stress profile can consist of regions of compressive and tensile stresses, following the shape of the initial profile (see Figs. \ref{fig:linear2}, \ref{fig:quadratic2}, \ref{fig:exp2}). For example, stress profiles with compressive-to-tensile transition were experimentally derived in Cu thin films \cite{Cancellieri2021}. Thus, the results of the present study show that such a residual stress profile is equilibrium and do not arise due to the limitation of the plasticity kinetics.

The number of dislocations to relax linear and exponential initial stress profiles appeared to be quite large (Figs. \ref{fig:linear1}, \ref{fig:linear2}, \ref{fig:exp1}, \ref{fig:exp2}). We can assume that it is less probable to proceed with a stress relaxation when many dislocations are needed to reach an equilibrium state. The emission of dislocations and their slip are kinetically limited, so an equilibrium state can be unavailable within a reasonable amount of time after the end of deposition. Moreover, such a large number of dislocations distributed along the film with a small thickness seems barely physical, as the dislocation cores would overlap, and we can no longer consider the problem using continuum elasticity theory.

Undoubtedly, the derived model being quite simple in its nature has limitations. First, the model considers only a single pile-up within a confined volume of width $d$. Generally speaking, in a grain, there is a set of neighboring dislocation pile-ups with a certain spacing. Each individual dislocation in a pile-up is affected by the forces induced not only by dislocations in that particular pile-up, but also by dislocations in the neighboring pile-ups. These cross-interactions define the spacing between the pile-ups. Taking these interactions into account, the final equation becomes more complicated. A rough estimate of $d$, the width of the film segment, can be made considering the mean spacing between dislocations derived by conventional models of single dislocation arrays, e.g., presented Freund and Suresh in Ref.\cite{Freund2004} (Chapter 7). For example, the simple model of the mean strain can be used:

\begin{equation}
    \frac{b_z}{8\pi L} \text{ln} \frac{4L}{b_z}=\varepsilon_{zy}^0-\frac{b_z}{p}
\end{equation}

Considering $L \approx 250\: \text{nm}$, $b_z \approx 0.5 \: \text{nm}$ and $\varepsilon_{zy}^0 \approx 1\, \text{GPa}/\mu$ with $\mu \approx 60 \; \text{GPa}$, the order of the film segment width is $d \approx 31\: \text{nm}$, so $k \approx 16$.
 
Second, in the present model, we suppose that stress relaxation begins only when the deposition is finished. It is generally not true, and stress relaxation is a continuous process that takes place upon deposition. Thus, it is a big question, which form of the initial stress distribution is proper when the deposition is finished. Moreover, plasticity upon growth will result in the formation of early dislocation pile-ups, so additional dislocation interactions must be considered in the grown film. The present simple model could be accurate only when the film grows much faster than the stress is being relaxed, i.e., the kinetics of relaxation mechanisms are negligible. Thus, we can say that the state of the film is frozen until the deposition is stopped completely. 

Third, a slip plane can be inclined to the film surface. For example, in fcc films, the angles of inclination between the (111) slip plane and the growth directions of [110] and [100] are 35.26$\degree$ and 54.73$\degree$, respectively. To incline the slip plane within the considered model (see Fig. \ref{fig:scheme}), we need to add the force due to the dislocation self-stress tensor component $\sigma_{xz}^{res}(x)$ to the force balance equation (Eq. \ref{eq7}). Thus, the solution for $b(\eta)$ needs to be re-derived, i.e., the updated integral equation in the form of Eq. \ref{eq9} must be solved.

In addition, this model is applicable only to thin films (or individual grains in a film) but not to multilayers. In multilayers, plasticity could not be described simply by glide of individual straight dislocations across the film thickness. Generally, plastic deformation in multilayers is related to the glide of dislocation loops within each individual layer, accompanied by cross-slipping at the interfaces between adjacent layers (see, e.g., Ref.\cite{Li2012b}). Thus, more complicated models are necessary to describe residual stress profiles in multilayer films.

\section{Conclusions}

In this study, we present an approach to model the residual stress gradient in a segment of a thin film based on dislocation pile-up theory. The basic equation was derived, which relates the derivative of the residual stress to the density distribution of dislocations. Referring to the classical work of Kuang and Mura, we derived the general expression of the dislocation distribution as a function of an arbitrary stress profile. Knowing the dislocation density distribution, we established an integro-differential equation, the solution of which is the residual stress distribution in a film segment. The residual stress profiles and the dislocation density distribution were then numerically determined for constant, linear, parabolic and exponential initial stress distributions. As intuitively expected, residual stress profiles tend to zero stress with an increasing thickness-to-width ratio of the film segment. In each case, dislocation pile-ups consist of dislocations with opposite Burgers vectors. The number of dislocations with positive and negative Burgers vectors depends on the initial distribution of stress in a film.

Non-uniform plasticity in the post-deposition stage is a key to the formation of residual stress distribution. This simple model is a crucial step toward understanding and modelling residual stress profiles in thin films.

\begin{acknowledgments}
This study was carried out according to the state task of Ioffe Institute.
\end{acknowledgments}

\bibliographystyle{aipnum4-1} 
\bibliography{literature}

\end{document}